\documentclass[english,journal = jpccck]{achemso}
\usepackage[T1]{fontenc}
\usepackage[latin9]{inputenc}
\usepackage{textcomp}
\usepackage{graphicx}

\makeatletter

\title{The rise of Single-Atom Catalysts}

\author{Deepak Kumar Rai }

\affiliation{Department of Chemistry, Southern University of Science and Technology,
Shenzhen 518000, China}

\email{dkriitb@gmail.com}

\makeatother

\usepackage{babel}
\begin{document}
\begin{abstract}
{\normalsize{}In recent years, single-atom catalysts attracted lots
of attention because of their high catalytic activity, selectivity,
stability, maximum atom utilization, exceptional performance, and
low cost. Single-atom catalyst contains isolated individual atom which
are coordinated with the surface atoms of support such as a metal
oxide or 2d - materials. In this review article, we present the advancement
in single-atom catalysis in recent years with a focus on the various
synthesis methods and their application in catalytic reactions. We
also demonstrate the reaction mechanism of a single-atom catalyst
for different catalytic reactions from theoretical aspects using density
functional theory.}{\normalsize\par}
\end{abstract}

\section{Introduction}

\label{sec:intro} 

In today's modern society, catalysts are widely used in the industrial
sectors such as petroleum refining, fuel cells, chemical intermediates,
pharmaceuticals, reduction of emission, and agrochemicals to increase
the reaction rate of the desired chemical reaction\cite{zhu2015engineering,liu2017noble,cai2018alkaline,li2004room,furstner2009gold,zheng2018preface,zhang2015catalysis}.
The conventional supported heterogeneous catalysts contain clusters
or nanoparticles dispersed on the surfaces of appropriate support
(i.e., metal oxides, 2d-materials, or porous metal-organic frameworks
nanomaterials). The atom utilization and selectivity of conventional
heterogeneous catalysts are very less, as only a part with a suitable
size of clusters or nanoparticles participates in the catalytic reaction.
Moreover, the remaining portion of clusters or nanoparticles does
not participate in the catalytic reaction, and it is not useful, may
be involved in unwanted reactions. The heterogeneous catalysts involved
in petroleum refining, new energy application, emission reduction,
and pharmaceuticals contain noble metal atoms like Pt, Pd, Ru, Rh,
Ir, Ag, and Au. These noble metals are costly and low abundant in
nature because of that these catalyst does not meet the current increasing
demand of industries, resulting minimization of the use of such catalyst
alter the catalytic activity of chemical reaction\cite{herzing2008identification,turner2008selective}.
To overcome these issues, researchers have found the most promising
way to increase the atom utilization and selectivity of catalysts
by reducing the size of nanoclusters to isolated individual atoms,
resulting in a catalyst containing a single atom on the surface of
a support. 

Single-atom catalysts (SACs) is a new class of catalyst which contain
isolated individual isolated atom dispersed or coordinated with the
surface atom of support. It exhibits high catalytic activity, stability,
selectivity, and 100 \% atom utilization because the whole surface
area of a single atom is exposed to reactants in catalytic reactions\cite{liang2015power,wang2018heterogeneous,yang2013single,liu2017catalysis,cheng2019single,wang2019single}.
In 2011, Zhang and co-workers\cite{Pt/FeOx} were the first to synthesized
and investigate experimentally and theoretically the catalytic activity,
selectivity and stability of single-atom Pt$_{1}$/FeO$_{x}$ catalyst
for CO oxidation. After that, It has attracted a lot of researchers,
and numerous SACs have been synthesized and developed in recent years.
By combing different noble atoms with different supports such as metal
oxides, 2d-materials, or porous metal-organic frameworks (MOFs) nanomaterials.
The SACs fabricated on different supports such as on metal oxides,
2d-materials, MOFs are {[}Pt$_{1}$/FeO$_{x}$\cite{Pt/FeOx}, Rh/ZrO$_{2}$\cite{Rh1/ZrO2},
Pt/$\theta$-Al$_{2}$O$_{3}$\cite{Pt/Al2O3}, Ir$_{1}$/FeO$_{x}$\cite{Ir1/FeOx},
Au$_{1}$/CeO$_{2}$\cite{Au/CeO2}, Au$_{1}$/Co$_{3}$O$_{4}$\cite{Au1/Co3O4},
Au$_{1}$/FeO$_{x}$\cite{Au1/FeOx}, Pd/FeO$_{x}$\cite{Pd/FeOx}
, Pd$_{1}$/TiO$_{2}$\cite{Pd/TiO2_liu2016photochemical}{]}, {[}
Pt/g-C$_{3}$N$_{4}$\cite{Pt-G-C3N4}, Pt/MoS$_{2}$\cite{Pt/MOS2_li2018synergetic},
Pt/GNS\cite{ALD_Sun}, Pd$_{1}$/graphene\cite{Pd_graphene_ALD_yan}{]},
{[} Co-SAs/N-C\cite{BMOF_yin2016single}, Fe-SAs/N-C\cite{Fe_ZIF-8_chen2017isolated},
Ni-SAs/N-C\cite{Ni-ZIF-8_zhao2017ionic}, and Ru-SAs/N-C\cite{MOF_wang2017uncoordinated}{]},
respectively. It has emerged as a new frontier in catalysis science
because of its excellent performance. 

In recent years, many researchers have reported that SACs shows excellent
performance in various catalytic reactions, such as CO oxidation\cite{Pt/FeOx,Ir1/FeOx_CO_liang2014theoretical,Ni1/FeOx_CO_liang2016theoretical,Au1/FeOx,Pt/Al2O3,M1/FeOx_li2014exploration,Pt/CeO2_CO_oxidation_ALD_wang},
water\textminus gas shift (WGS)\cite{Ir1/FeOx,WGS_review_flytzani2012atomically,wgs_thomas2011can,WGS_Au_flytzani2013gold,Au/CeO2_WGS_DFTsong2014mechanistic,Au-OHx/TiO2_WGSyang2013atomically,liang2020dual},
water splitting, hydrogenation reaction, carbon dioxide reduction,
etc. Despite the excellent performance of SACs, it has some limitations
and disadvantages. The stabilization of single atoms on the surfaces
of support is a very challenging process due to the agglomeration
of single-atoms. It needs advanced techniques for synthesis, which
we have discussed in the next section.

Remainder of the paper is organized as follows. In the next section
\ref{sec:Synthesis-of-SAC}II, we briefly discuss the advanced synthesis
methods of SACs, while in section \ref{sec:Synthesis-of-SAC}III,
we describe the application of SACs for different chemical reactions
and their reaction mechanism from theoretical aspects. Finally in
section \ref{sec:Summary-and-Conclusions}IV, we summarize our review
article.

\section{\label{sec:Synthesis-of-SAC}Synthesis of Single Atom Catalysis }

In this section, we present the various synthesis methods for the
fabrication of single-atom catalysts. The stabilization of single
atom on the surfaces of metal oxide or two-dimensional materials is
a very challenging process due to the agglomeration of single-atoms
and the tendency to form nanoparticles and clusters on the surfaces.
The agglomeration of single atoms happens because the surface energy
of nanoparticles and clusters is less than single-atom. So, advanced
synthesis methods such as impregnation method, co-precipitation method,
other-wet-chemical synthesis method, atomic layer deposition method,
and metal-organic frameworks derived method are used for fabrication
single-atom catalysts , are discussed below.

\subsection{Impregnation Method}

\begin{figure}[H]
\begin{centering}
\includegraphics[scale=0.4]{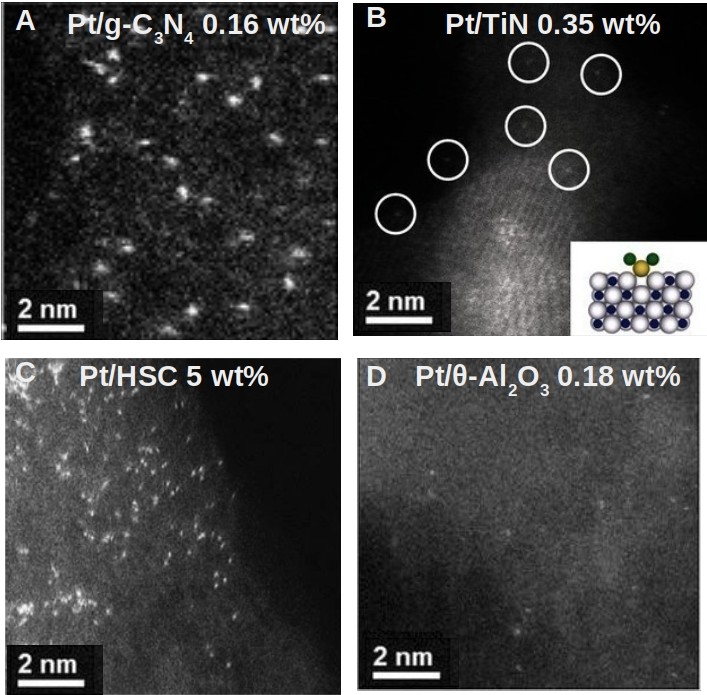}
\par\end{centering}
\caption{\label{fig:impregnation_HAADF-STEM} The figure represents the HAADF-STEM
images of different single atom catalysts synthesized using impregnation
method: (a) Pt/g-C$_{3}$N$_{4}$, (b) Pt/TiN, (c) Pt/LSC, (d) Pt/$\theta-$Al$_{2}$O$_{3}$.
(a) reprinted/reproduce from Ref. {[}25{]} with the permission of
Wiley-VCH publishing group, copyright 2016, (b) reprinted/reproduce
from Ref. {[}43{]} with the permission of Wiley-VCH publishing group,
copyright 2016, (c) reprinted/reproduce from Ref. {[}44{]} with the
permission of Nature publishing group, copyright 2016, (d) reprinted/reproduce
from Ref. {[}18{]} with the permission of American Chemical society
publishing group, copyright 2013.}
\end{figure}

For the synthesis of the single-atom or supported catalyst impregnation
method is the simplest and economical method. In this method, a small
amount of solution containing active metal precursor is mixed with
catalyst support, and using the ion-exchange and adsorption process
active metals stabilized on the surface of support. Li et al.\cite{Pt-G-C3N4}
synthesized Pt/g-C$_{3}$N$_{4}$ (see Fig. \ref{fig:impregnation_HAADF-STEM}
A) by performing liquid phase reaction between graphitic carbon nitride
(g-C$_{3}$N$_{4}$) and H$_{2}$PtCl$_{6}$, followed by annealing
at low temperature, and this catalyst exhibit high activity for H$_{2}$
evolution. They prepared four samples of supported catalyst with different
weight percentage (i.e. 0.075\%, 0.11\%, 0.16\%, 0.38\%) of metal
loading, and found that at 0.16 wt\% a bright spot center of Pt atoms
are distributed on the surface on the surface of g-C$_{3}$N$_{4}$,
can be seen in HAAD-STEM images. When the weight percentage is increased,
up to 0.38\% aggregation of small Pt atoms is observed on the surface.
Yang et al. \cite{Pt/TiN} prepared Pt/TiN catalyst (see Fig. \ref{fig:impregnation_HAADF-STEM}
B) using wetness impregnation method, in which a small amount (0.35
wt\%) of Pt atoms is loaded on the surface of acid-treated TiN support,
and this catalyst is found to be active for oxygen reduction reactions,
formic acid, and methanol oxidation.

Yang et al. observed the formation of Pt nanoparticles on the surface
of support if the weight percentage is increased above 0.35\%. Recently,
Choi et al.\cite{choi2016tuning} synthesized highly loaded (5 wt\%)
Pt/S-ZTC catalyst (see Fig. \ref{fig:impregnation_HAADF-STEM} C)
using wet-impregnation method, in which Pt atom is atomically dispersed
on the surface of sulfur-doped zeolite-templated carbon (S-ZTC). The
doped sulfur and unique three-dimensional structure of ZTC stabilize
the loaded Pt atoms on the support surface. They have reported Pt/S-ZTC
exhibit high activity for oxygen reduction reaction. Kwon et al.\cite{Rh1/ZrO2}
have studied the activation of methane for methanol production using
Rh/ZrO$_{2}$SACs, which the prepared by wet impregnation method.

Moses-DeBusk et al.\cite{Pt/Al2O3} have studied the CO oxidation
activity of a single Pt atom supported on $\theta$-alumina ($\theta$-Al$_{2}$O$_{3}$).
They have synthesized Pt/$\theta$-Al$_{2}$O$_{3}$ SAC (see Fig.
\ref{fig:impregnation_HAADF-STEM} D) by mixing alumina in an aqueous
solution of chloroplatinic acid, heated at mild temperature for 30
hours, and placed on a rotovap for water evaporation. Resulting free
flow powder is kept in an alumina crucible, and pyrolysis is done
with elevated temperature 1 $^{\deg}$C/min to 450 $^{\circ}$C for
4 hours for obtaining Pt/$\theta$-Al$_{2}$O$_{3}$ SAC. The HAADF-STEM
images of single atoms catalysts Pt/g-C$_{3}$N$_{4}$\cite{Pt-G-C3N4},
Pt/TiN\cite{Pt/TiN}, Pt/LSC\cite{choi2016tuning} and Pt/$\theta-$Al$_{2}$O$_{3}$\cite{Pt/Al2O3}
with 0.16 wt\%, 0.35 wt\%, 5wt\% and 0.18 wt\%, respectively, are
presented in Fig. \ref{fig:impregnation_HAADF-STEM}. 

It is challenging to produce uniformly distributed and highly loaded
SACs with this method because it depends on the ability of support
to adsorb the metal atoms, .i.e, the loading and distribution depends
on the number of anchoring sites present on the surface of support. 

\subsection{Co-precipitation Method }

\begin{figure}[H]
\begin{centering}
\includegraphics[scale=0.39]{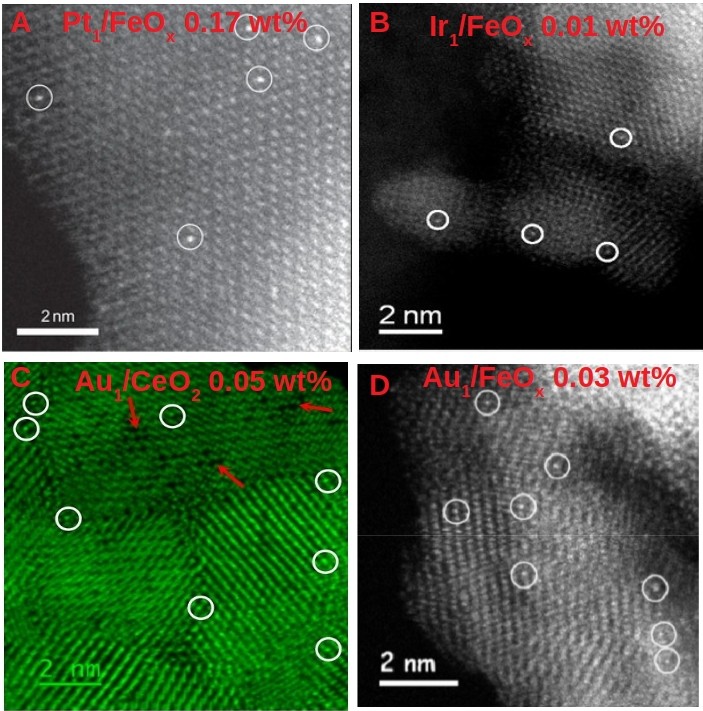}
\par\end{centering}
\caption{\label{fig:Coprecipitation-HAAD-STEM}The figure represents the HAADF-STEM
images of different single atom catalyst synthesized using co-precipitation
method: (a) Pt$_{1}$/FeO$_{x}$, (b) Ir$_{1}$/FeO$_{x}$, (c) Au$_{1}$/CeO$_{2}$,
(d) Au$_{1}$/FeO$_{x}$. (a) reprinted/reproduce from Ref. {[}16{]}
with the permission of Nature publishing group, copyright 2011, (b)
reprinted/reproduce from Ref. {[}19{]} with the permission of American
Chemical society publishing group, copyright 2013, (c) reprinted/reproduce
from Ref. {[}20{]} with the permission of American Chemical society
publishing group, copyright 2015, (d) reprinted/reproduce from Ref.
{[}22{]} with the permission of Tsinghua university press and Springer
publishing group, copyright 2015.}
\end{figure}

Co-precipitation is a convenient, cost-effective, and less time-consuming
method for the synthesis of nanoparticles. This method is slightly
different from the impregnation method; here, metals atom are incorporated
in the interstitial sites of support, not distributed on the surface
of the support. In this method, anionic and cationic solution are
mixed and simultaneously nucleation, growth, coarsening, and/or agglomeration
processes starts. After agglomeration, we have to followed three more
steps, i.e., precipitation, filtration, and calcination, and finally,
nanoparticle is obtained. Recently, Zhang's research group have reported
that they were the first one to fabricate SAC containing isolated
Pt atoms uniformly dispersed on the iron oxide (FeO$_{x}$) support
using the co-precipitation method\cite{Pt/FeOx,Pt/FeOx_No_reduction}.
Two samples of Pt$_{1}$/FeO$_{x}$ (see Fig. \ref{fig:Coprecipitation-HAAD-STEM}
A) were prepared, with 0.17 wt\% and 2.5 wt\%, using an aqueous solution
of chloroplatinic acid (H$_{2}$PtCl$_{6}$.6H$_{2}$O) and ferric
nitrate Fe(NO$_{3}$)$_{3}$.9H$_{2}$O with precipitation agent sodium
carbonate (Na$_{2}$CO$_{3}$) at 50 $^{\circ}$C, and the PH value
is maintained around 8. The resulting sample was dried at 60 $^{\circ}$C
for five hours and calcined at 400 $^{\circ}$C for five hours. Furthermore,
samples were reduced at 200 $^{\circ}$C for half an hour with 10\%
H$_{2}$/He flow rate. They also reported that at low Pt loading 0.17
wt\%, uniformly dispersed isolated Pt atom on the FeO$_{x}$ support
can be seen HAADF images, whereas, at 2.5 wt\%, mixture of Pt atoms,
2D structure of Pt atoms and cluster of Pt atoms is observed. This
SAC shows excellent activity and stability for CO oxidation and NO
reduction. 

In addition to that Zhang and co-workers, using co-precipitation method,
have synthesized series of SACs such as Ir$_{1}$/FeO$_{x}$\cite{Ir1/FeOx}
(see Fig. \ref{fig:Coprecipitation-HAAD-STEM} B), Au$_{1}$/CeO$_{2}$\cite{Au/CeO2}
(see Fig. \ref{fig:Coprecipitation-HAAD-STEM} C), Au$_{1}$/Co$_{3}$O$_{4}$\cite{Au1/Co3O4},
Au$_{1}$/FeO$_{x}$\cite{Au1/FeOx} (see Fig. \ref{fig:Coprecipitation-HAAD-STEM}
D), and Pd/FeO$_{x}$\cite{Pd/FeOx}, which exhibits excellent activity
and stability for water-gas shift reactions and CO oxidation.

Xing et al.\cite{Pt/TiO2_Xin} have prepared single atom photo-catalyst
containing isolated metal atoms (Pt, Pd, Ru and Rh) uniformly dispersed
on titanium oxide (TiO$_{2}$) support using co-precipitation method
and studied their activity and stability for water-splitting reaction.
They have synthesized 4 samples for Pt/TiO$_{2}$ with different metal
loading percentage 0.2 wt\%, 0.5 wt\%, 2wt\% and 1Pt/TiO$_{2}$(PD)
(pure photo deposited 1 wt\% of Pt nanoparticles), and found that
H$_{2}$ evolution rate for 0.2-Pt/TiO$_{2}$ is 169.6 $\mu$mol/h,
which is 23, 57, and 136 times more than 1Pt/TiO$_{2}$(PD), 0.5-Pt/TiO$_{2}$
and 2-Pt/TiO$_{2}$, respectively. The H$_{2}$ evolution rate for
Pd, Ru, and Rh nanoparticles supported on TiO$_{2}$ is 7, 7 and 13
times less than 0.2-Pt/TiO$_{2}$, respectively. The HAADF-STEM images
of single atoms catalysts Pt$_{1}$/FeO$_{x}$\cite{Pt/FeOx}, Ir$_{1}$/FeO$_{x}$\cite{Ir1/FeOx},
Au$_{1}$/CeO$_{2}$\cite{Au/CeO2}, and Au$_{1}$/FeO$_{x}$\cite{Au1/FeOx}
with 0.17 wt\%, 0.01 wt\%, 0.05 wt\% and 0.03 wt\%, respectively,
are presented in Fig. \ref{fig:Coprecipitation-HAAD-STEM}. 

The advantages to this method are; it is a simple, rapid, easy to
control the particle size and composition of the final product, energy-efficient
and does not need organic solvent. Moreover, disadvantages of this
method are; it does not apply to uncharged species, trace of impurities
also get precipitated, reproducibility problem and does not work well
if the reactants have very different precipitation rate.

\subsection{Other Wet-Chemical Synthesis Method}

\begin{figure}[H]
\begin{centering}
\includegraphics[scale=0.35]{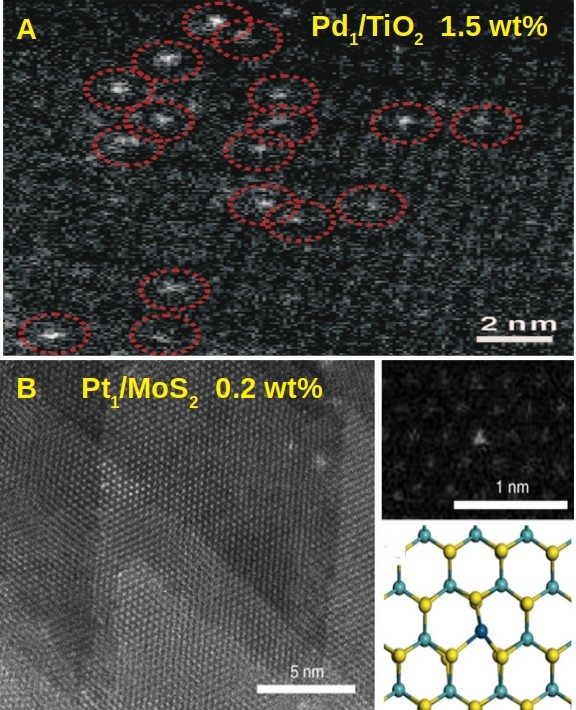}
\par\end{centering}
\caption{\label{fig:wet-chemical-synthesis}The figure represents the HAADF-STEM
images of different single atom catalyst synthesized using other wet-chemical
method: (a) Pd$_{1}$/TiO$_{2}$, (b) Pt$_{1}$/MoS$_{2}$. (a) reprinted/reproduce
from Ref. {[}24{]} with the permission of Science publishing group,
copyright 2016, (b) reprinted/reproduce from Ref. {[}26{]} with the
permission of Macmillan Publishers Limited and part of Springer Nature
publishing group, copyright 2018.}
\end{figure}

Impregnation and Co-precipitation are the traditional wet-chemical
synthesis method, but Liu et al.\cite{Pd/TiO2_liu2016photochemical}
have used unique wet-chemical synthesis methods for the fabrication
of single atom Pd$_{1}$/TiO$_{2}$ catalyst (see Fig. \ref{fig:wet-chemical-synthesis}
A) with a high metal loading up to 1.5 wt\%. They dispersed a solution
of H$_{2}$PdCl$_{4}$ on the surface of TiO$_{2}$ support, the resulting
mixture is exposed to UV rays for 10 min. After that, the irradiated
sample is washed with water, and a single atom Pd$_{1}$ /TiO$_{2}$
catalyst is obtained. Form transmission electron microscopy (TEM)
images and extended x-ray absorption fine structure (EXAFS) spectra,
it is concluded that the formation of Pd clusters or nanoparticles
are not observed. This catalyst exhibits very high catalytic activity
and stability for the hydrogenation of C=C and C=O.

Recently, Li et al.\cite{Pt/MOS2_li2018synergetic} synthesized single
atom Pt/MoS$_{2}$ catalyst (see Fig. \ref{fig:wet-chemical-synthesis}
B) by injecting solution of K$_{2}$PtCl$_{6}$ using syringe pump
into the mixture of MoS$_{2}$ nanosheets, ethanol, and water. During
the chemisorption process, Mo atoms are replaced by Pt atoms in MoS$_{2}$
nanosheets. Pt/MoS$_{2}$ catalysts with different Pt loading percentages
0.2, 1.0, 5.0, 7.5 are synthesized by changing the concentration of
K$_{2}$PtCl$_{6}$, and EXAFS spectra of all these catalysts confirmed
that only isolated Pt is present on the surface of MoS$_{2}$. Researchers
also investigated its catalytic activity for the conversion of CO$_{2}$
into methanol without the formation of formic acid. The HAADF-STEM
images of single atoms catalysts Pd$_{1}$/TiO$_{2}$\cite{Pd/TiO2_liu2016photochemical},
and Pt$_{1}$/MoS$_{2}$\cite{Pt/MOS2_li2018synergetic} with 1.5
wt\%, and 0.2 wt\%, respectively, are presented in Fig. \ref{fig:wet-chemical-synthesis}. 

\subsection{Atomic Layer Depositions Method}

Atomic layer deposition (ALD) is a subclass of chemical vapor deposition,
attracting many researchers because of its ability to deposit noble
metal atoms and their oxides uniformly with a desirable thickness
on the substrate by using using sequential and self limitig surface
reaction\cite{ALD_Cheng,ALD_george,ALD_neill,ALD_Lu,ALD_Liu}. Generally,
In this method, two precursors are used, and the deposition process
involves four steps\cite{ALD_cheng_nano_energy}; (1) Initially, precursor
is inserted in the chamber and allowed to react with the substrate;
(2) Purging of reaction chamber by use of carrier gas; (3) Second
precursor is inserted in the reaction chamber and allowed to react
with substrate containing first precursor; (4) At last, purging of
reaction chamber is done, and sample is obtained. By repeating the
cycles, a desired thickness of the precursor can be deposited. 

\begin{figure}[H]
\begin{centering}
\includegraphics[scale=0.4]{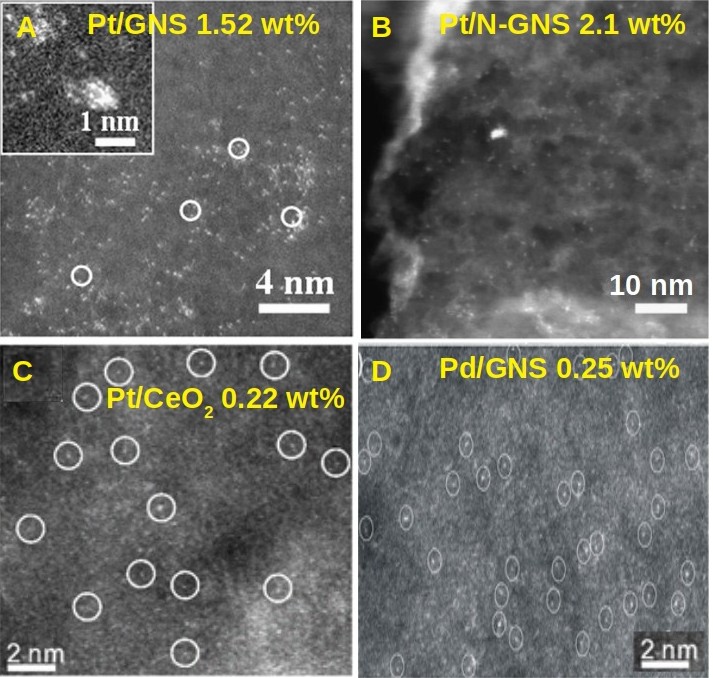}
\par\end{centering}
\caption{\label{fig:ALD_HAAD_STEM}The figure represents the HAADF-STEM images
of different single atom catalyst synthesized using atomic layer deposition
method: (a) Pt/GNS, (b) Pt/N-GNS, (c) Pt$_{1}$/CeO$_{2}$, (d) Pd/GNS.
(a) reprinted/reproduce from Ref. {[}27{]} with the permission of
Nature publishing group, copyright 2013, (b) reprinted/reproduce from
Ref. {[}53{]} with the permission of Nature publishing group, copyright
2016, (c) reprinted/reproduce from Ref. {[}36{]} with the permission
of American Chemical society publishing group, copyright 2016, (d)
reprinted/reproduce from Ref. {[}28{]} with the permission of American
Chemical society publishing group, copyright 2015.}
\end{figure}

Sun et al.\cite{ALD_Sun} synthesized heterogeneous catalysts consists
of isolated Pt atoms, Pt-clusters, Pt-nanoparticles dispersed on the
surface graphene nanosheets (GNS) using ALD method, and also reported
that these novel catalyst shows remarkable catalytic activity for
methanol oxidation, almost ten times higher than the commercial carbon
supported Pt (Pt/C) catalyst. For the synthesis of Pt/GNS catalyst
(methylcyclopentadienly)-trimethylplatinum (MeCpPtMe$_{3}$, 98\%
purity) and oxygen (99.9995\%) used as precursors and nitrogen (99.9995\%)
use as purge and carrier gas. The HAAD-STEM images of Pt/GNS catalyst
synthesized with 50, 100, and 150 ALD cycles, reveals that isolated
pt atoms and small cluster (<1 nm) are present in the 50ALD-Pt/GNS
(see Fig. \ref{fig:ALD_HAAD_STEM} A), whereas in the 100ALD-Pt/GNS,
and 150ALD-Pt/GNS the size cluster approaches to 2 nm and 4 nm, respectively.
Recently, Cheng et al.\cite{Pt/N-GNS_ALD_HER_cheng} synthesized Pt/N-GNS
SAC by same ALD technique discussed above, in which isolated Pt atoms
are uniformly dispersed on the surface of nitrogen-doped graphene
nano-sheets, and also investigated its activity for Hydrogen evolution
reaction. They also reported, Pt/NGNs exhibits enhanced catalytic
activity ($\approx$ 37 times more than Pt/C) and high stability.
The Pt loading of 2.1 and 7.6 wt\% for 50 and 100 ALD cycles, respectively,
was confirmed by inductively coupled plasma atomic emission spectroscopy.
Similarly, as above, a bright spot of isolated Pt atoms, as well as
tiny Pt cluster, are observed in 50ALD-Pt/NGNs (see Fig. \ref{fig:ALD_HAAD_STEM}
B), whereas in 100ALD-Pt/NGNs, the size of Pt clusters becomes larger
and formation of nanoparticles, as well as new cluster, is observed.

Lu and co-workers\cite{Pd_graphene_ALD_yan} used ALD technique for
preparation of single atom Pd1/graphene catalyst (see Fig. \ref{fig:ALD_HAAD_STEM}
D), palladium hexafluoroacetylacetate (Pd(hfac)$_{2}$, Sigma Aldrich,
99\%) and formalin (Aldrich, 37\% HCHO and 15\% CH$_{3}$OH in aqueous
solution) used as precursors and N$_{2}$ (99.999\%, purity) as carrier
and purge gas. Researchers explored the hydrogenation of 1,3-butadiene
using Pd$_{1}$/graphene SAC and observed excellent durability for
catalytic deactivation and remarkable catalytic performance, i.e.,
100\% butenes selectivity and 95\% conversion at 50 $^{\circ}$C.
Wang et al.\cite{Pt/CeO2_CO_oxidation_ALD_wang} from Lu group have
synthesized single Pt$_{1}$/CeO$_{2}$ (see Fig. \ref{fig:ALD_HAAD_STEM}
C) catalyst and studied its activity in water promoted CO oxidation
and reported that the contribution of water in production CO2 using
Pt$_{1}$/CeO$_{2}$is 50\% via a water-mediated Mars-Van Krevelen
(MvK) mechanism. 

Piernavieja-Hermida et al.\cite{Pd/Al2O3_TiO2_ALD_Piernavieja-Hermida}
developed an exciting way to stabilized single Pd atom on the surface
of Al$_{2}$O$_{3}$ by depositing an ultra-thin layer of TiO$_{2}$
protective coatings. First, Pd(hfac)$_{2}$precursor is allowed to
chemisorbed on the surface of Al$_{2}$O$_{3}$ using ALD; after that,
TiO$_{2}$ is deposited on the substrate using tetrachloride and ionized
water. The TiO$_{2}$ selectively grows on the substrate, not on the
Pd(hfac)$_{2}$ because of the presence of remaining (hfac)$_{2}$,
which prevents its growth on Pd. The massive structure of (hfac)$_{2}$
forms a nanocavity around the Pd atoms of same the dimension, and
at last, these ligands are removed using formalin (HCHO) for obtaining
the TiO$_{2}$ protected Pd/Al$_{2}$O$_{3}$catalyst. They also reported
that the thermal stability of this catalyst significantly increased
because of the nanocavity thin-film structure. The HAADF-STEM images
of single atoms catalysts Pt/GNS\cite{ALD_Sun}, Pt/N-GNS\cite{Pt/N-GNS_ALD_HER_cheng},
Pt$_{1}$/CeO$_{2}$\cite{Pt/CeO2_CO_oxidation_ALD_wang}, and Pd/GNS\cite{Pd_graphene_ALD_yan}
with 1.52 wt\%, 2.1 wt\%, 0.22 wt\% and 0.25 wt\%, respectively, are
presented in Fig. \ref{fig:ALD_HAAD_STEM}. 

The major disadvantages of the ALD technique are; it is a time-consuming
method and ALD instruments and running cost are very expensive. 

\subsection{Metal-Organic Frameworks Derived Method}

Metal-organic frameworks (MOFs)\cite{MOF_zhang2016efficient,MOF_wang2017uncoordinated,MOF_he2018zirconium}
are the porous compound in which of metal ions or clusters are attach
with organic ligands to form 1-, 2- or 3- dimensional structure, and
could be used as precursors or as support in the synthesis of SACs.
Unique characteristics of MOFs such as high surface area, ordered
pore structure with uniform sizes makes them ideal substrate for loading
of single atom. In the synthesis of SACs, MOFs are emerging as a new
research frontier because of the following reasons; (1) Tunable pore
size enables MOFs to encapsulate metal precursors and prevent form
the agglomeration. (2) The high surface area of MOFs provides a large
number of anchoring sites for dispersion of metal precursors. (3)
A variety of organic ligands serve as active anchoring sites for various
precursors. (4) Using the pyrolysis method, various MOFs can be easily
converted into N-doped carbon materials, and act as a ideal substrate
for dispersion and stabilization of metal precursors.

\begin{figure}[H]
\begin{centering}
\includegraphics[scale=0.4]{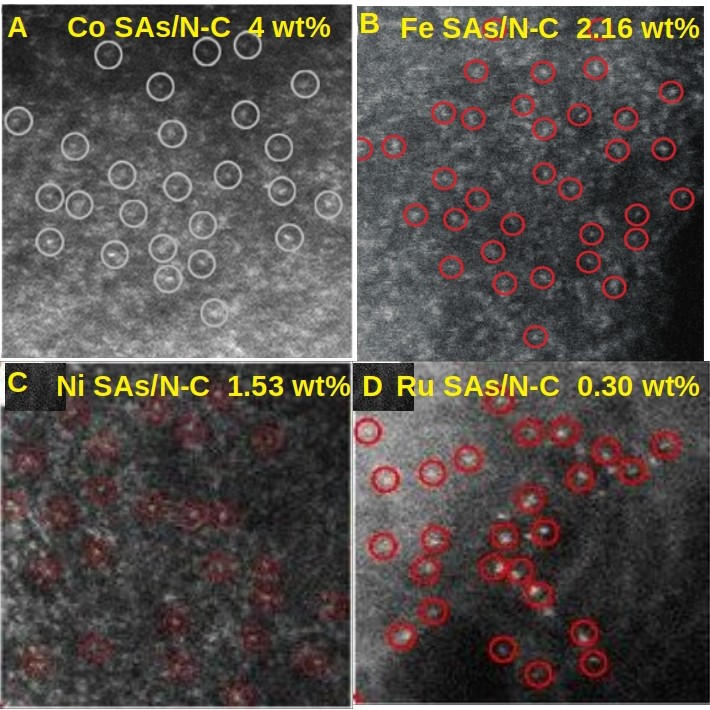}
\par\end{centering}
\caption{\label{fig:MOFs_HAAD_STEM-1}The figure represents the HAADF-STEM
images of different single atom catalyst synthesized using metal-organic
frameworks method: (a) Co SAs/N-C, (b) Fe SAs/N-C, (c) Ni SAs/N-C,
(d) Ru SAs/N-C. (a) reprinted/reproduce from Ref. {[}29{]} with the
permission of Wiley-VCH publishing group, copyright 2016, (b) reprinted/reproduce
from Ref. {[}30{]} with the permission of Wiley-VCH publishing group,
copyright 2017, (c) reprinted/reproduce from Ref. {[}31{]} with the
permission of American Chemical society publishing group, copyright
2017, (d) reprinted/reproduce from Ref. {[}32{]} with the permission
of American Chemical society publishing group, copyright 2017.}
\end{figure}

Yin et al.\cite{BMOF_yin2016single} from Li group synthesized single
atom Co/N-C catalyst (see Fig. \ref{fig:MOFs_HAAD_STEM-1} A) with
high metal loading up to 4 wt\%, which consists Co atom dispersed
on the surface of nitrogen-doped porous carbon and investigated its
activity for oxygen reduction reaction. By performing the pyrolysis
of bimetallic Zn/Co metal-organic framework (BMOF) at 800 $^{\circ}$C,
the Co and Zn ions are reduced by carbonization of organic ligands
and further evaporation of Zn takes place because of its low boiling
point and single atom Co/N-C catalyst is obtained. Wang et al.\cite{BMOF_wang2018regulation}
from the same group reported that the coordination number of Co atom
could be controlled by changing the pyrolysis temperature, for example
they fabricated three single atom Co-N$_{4}$, Co-N$_{3}$ and Co-N$_{2}$
catalyst by keeping the pyrolysis temperature at 800, 900 and 100
$^{\circ}$C, respectively. 

Chen et al.\cite{Fe_ZIF-8_chen2017isolated} also from Li group synthesized
isolated Fe atom supported on nitrogen-doped porous carbon (Fe SAs/N-C)
catalyst (see Fig. \ref{fig:MOFs_HAAD_STEM-1} B) with metal loading
up to 2.16 wt\% and reported its excellent activity for oxygen reduction
reaction compared to Pt/C and most non-expensive-metal catalyst. They
mixed Fe(acac)$_{3}$ and zeolitic imidazolate frameworks (ZIF-8)
and used encapsulated-precursor pyrolysis technique for the synthesis
of Fe/N-C catalyst. The molecular-scale cage structure of ZIF-8 formed
by assembly of Zn$^{2+}$ and 2-methylimidazole traps one Fe(acac)$_{3}$
molecule. After that, pyrolysis of resulting mixture at 900 $^{\circ}$C
under Ar gas converts ZIF-8 into nitrogen-doped porous carbon, and
simultaneously Fe(acac)$_{3}$ was reduced by carbonized organic ligands,
and Fe SAs/N-C catalyst is obtained. 

Zhao et al.\cite{Ni-ZIF-8_zhao2017ionic} prepared isolated Ni atom
dispersed on the nitrogen-doped porous carbon (Ni SAs/N-C) (see Fig.
\ref{fig:MOFs_HAAD_STEM-1} C) with metal loading up to 1.53 wt\%
and investigated its activity for electroreduction of CO$_{2}$. The
homogeneous aqueous solution of Ni(NO$_{3}$)$_{2}$ was mixed with
a solution containing ZIF-8 powder dispersed in n-hexane and actively
stirred for 3 hours so that salt completely absorbed, resulting sample
was centrifuged and dried at 65 $^{\circ}$C for 6 hours. After that,
pyrolysis of the sample at 1000 $^{\circ}$C was done in the presence
of Ar gas, during which the ZIF-8 is converted into nitrogen-doped
porous carbon, simultaneously Zn atoms evaporate due to its low boiling
point, creating nitrogen-rich sites. These sites are occupied by Ni$^{2+}$,
and act as a fence and prevent Ni atom from agglomeration; finally,
Ni SAs/N-C catalyst is obtained. 

Wang $et$ $al.$\cite{MOF_wang2017uncoordinated} synthesized Ru
SAs/N-C catalyst (see Fig. \ref{fig:MOFs_HAAD_STEM-1} D), which contains
single Ru atom dispersed on the nitrogen-doped porous carbon with
metal loading percentage 0.30 wt\% and reported that it exhibits high
catalytic and selectivity for hydrogenation of quinolines. They used
amine derivative MOF UiO-66-NH$_{2}$ (Zr$_{6}$O$_{4}$(OH)$_{4}$(BDC)$_{6}$-NH$_{2}$)
for synthesizing of Ru SAs/N-C catalyst, first, they mixed RuCl$_{3}$,
ZrCl$_{4}$ and H$_{2}$BDC-NH$_{2}$ with an aqueous solution of
DMF and HAs. After that resulting mixture is centrifuged and washed
with methanol and DMF and then heated at 700 $^{\circ}$C for 3 hours
in the presence of Ar gas, a black powder containing nitrogen-doped
porous carbon (N-C) decorated with Ru and small ZrO$_{2}$ species
is obtained. The small ZrO$_{2}$ spices attach with N-C were etched
by adding HF solution, and Ru SAs/N-C is formed. 

Wei $et$ $al.$\cite{Pd_Pt_Au_nano_particles_ZIF-8_wei2018direct}
synthesized catalyst containing a single Pd atom supported on the
nitrogen-doped porous carbon from Pd-nanoparticles by employing the
top-down method, and also reported its excellent catalytic activity
and selectivity for semi-hydrogenation of acetylene to ethylene. First,
ZIF-8 nanocrystal was grown on the surface Pd-nanoparticles, by mixing
Pd-nanoparticles in an aqueous solution Zn(NO$_{3}$)$_{2}$ and 2-methylimidazole
solution. After that, the resulting mixture was heated at 900 $^{\circ}$C
in presences of inert gas for 3 hours, Pd-nanoparticles were transformed
in single atoms and distributed within the substrate, and meanwhile,
ZIF-8 is converted into nitrogen-doped porous carbon. Finally, single-atom
Pd SAs/N-C is obtained having a thermodynamically stable Pd-N$_{4}$
structure. Using the same technique, they have also synthesized Pt
SAs/N-C and Au SAs/N-C catalyst.

Using ZIF-8 MOF and pyrolysis method, Yang $et$ $al.$\cite{Ni-ZIF-8_nanoparticles_yang2018situ}
synthesized Ni SAs/N-C catalyst by transforming Ni nanoparticles into
Ni single atom, mostly dispersed on the surface of N-doped porous
carbon substrate and tested its activity and selectivity for electroreduction
of CO$_{2}$.

Recently, Zhang $et$ $al.$\cite{Fe/N-C_zhang2019single} prepared
Fe$_{1}$-N-C SAC using porphyrinic MOF (PCN-222), the catalyst contains
isolated Fe atom dispersed on the surface of nitrogen-doped porous
carbon substrate, and also reported its activity for nitrogen reduction
reaction is better than Co$_{1}$-N-C and Ni$_{1}$-N-C. Initially,
they synthesized Fe-TPPCOOMeCl (iron (III) meso-tetra(4-methoxycarbonylphenyl)
porphine chloride (Fe-TPPCOOMeCl) by dissolving TPPCOOMe and FeCl$_{2}$.4H$_{2}$O
in a DMF solution and heated for 6 hours at 160 $^{\circ}$C. Then
Fe-TCPP is obtained by mixing Fe-TPPCOOMeCl THF, MeOH and KOH, and
heated for 6 hours at 85 $^{\circ}$C. After that, Fe-TCPP, ZrOCl$_{2}$.8H$_{2}$O,
H$_{2}$-TCPP, DMF and CF$_{3}$COOH were mixed and heated for 18
hours at 120 $^{\circ}$C for the formation of PCN-222(Fe). At last
pyrolysis of PCN-222(Fe) sample is done at 800 $^{\circ}$C for 3
hours in the presence of N$_{2}$ gas, and Fe$_{1}$-N-C catalyst
is obtained. The HAADF-STEM images of single atoms catalysts Co SAs/N-C\cite{BMOF_yin2016single},
Fe SAs/N-C\cite{Fe_ZIF-8_chen2017isolated}, Ni SAs/N-C\cite{Ni-ZIF-8_zhao2017ionic},
and Ru SAs/N-C\cite{MOF_wang2017uncoordinated} with 4 wt\%, 2.16
wt\%, 1.53 wt\% and 0.30 wt\%, respectively, are presented in Fig.
\ref{fig:MOFs_HAAD_STEM-1}. 

\section{\label{sec:Application-of-SAC}Application of Single-Atom Catalysis}

In recent years, researchers have reported the synthesis and catalytic
behavior of many SACs. They found that these SACs show high catalytic
activity, selectivity, and stability because of the maximum utilization
of single atom (almost 100\% utilization) during reactions and strong
bonding between the single atom and the anchoring sites on the supported
surfaces. Therefore, the application of many SACs in different catalytic
reactions such as CO oxidation, water-gas shift reaction, water splitting
reaction, oxygen reduction reaction, methanol oxidation reaction,
C-H activation reactions, Hydrogen evolution reaction, carbon dioxide
reduction reaction and Hydrogenation reaction, are discussed below.

\subsection{CO Oxidation Reaction}

In the field of catalyst science, CO oxidation is one of the most
studied reaction because of its importance in protecting our environment
by purifying poisonous exhaust gases coming from motor vehicles and
various Industries\cite{CO_Oxidation_gardner1991comparison,CO_Oxidation_haruta1997size}.
Moreover, CO oxidation is the most crucial step in water-gas-shift
reaction\cite{WGS_gokhale2008mechanism,WGS_lin2013remarkable} and
in fuel cells application for eliminating CO from reforming gas. Zhang
and co-workers\cite{Pt/FeOx} were the first to investigate experimentally
and theoretically the catalytic activity, selectivity and stability
of single-atom Pt$_{1}$/FeO$_{x}$ catalyst for CO oxidation, and
relativistic density functional theory was used for theoretical investigation. 

\begin{figure}
\begin{centering}
\includegraphics[scale=0.5]{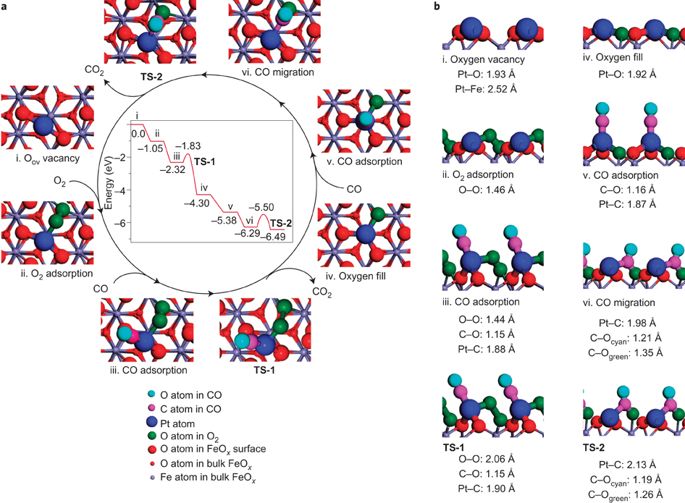}
\par\end{centering}
\caption{\label{fig:Co-oxidation-Pt-FeOx}(a) Top and (b) side view of proposed
reaction pathways for CO oxidation on Pt$_{1}$/FeO$_{x}$. The calculated
relative energy of proposed reaction pathways is presented in circle.
(reprinted/reproduce from Ref. {[}16{]} with the permission of Nature
publishing group, copyright 2011).}
\end{figure}

For computation, they used Fe- and O$_{3}$- terminated Fe$_{2}$O$_{3}$
(011) surfaces, and after optimization found that the most likeliest
position of Pt atom is 3-fold hollow sites on the O$_{3}$ terminated
surface, where Pt atom is linked with three oxygen atom. From HAAD
images, they observed that the single Pt atom exactly replaces the
single Fe atom located at 3-fold hollow sites of O$_{3}$-terminated
surface. Before testing the catalytic performance, the Pt$_{1}$/FeO$_{x}$
catalyst was reduced by flowing H$_{2}$/He gas for 30 min at 200
$^{\circ}$C.The oxidation of CO on the surface of Pt$_{1}$/FeO$_{x}$follows
Langmuir-Hinshelwood (H-L) mechanism and the step by step reaction
mechanism is shown in Fig. \ref{fig:Co-oxidation-Pt-FeOx}. After
prereduction by H$_{2}$, the oxygen vacancy (O$_{vac}$) near the
Pt atom was created by reducing the stoichiometric hematite surfaces
partially (step i), which provides an active site for adsorption of
O$_{2}$ molecule. A Similar theoretical model was designed for computation
by removing one oxygen atom, which is connected to the Pt atom, the
oxygen coordination number of Pt atom reduces from 3 to 2. In step
ii, O$_{2}$ molecule is adsorbed with adsorption energy 1.05 eV,
and optimize O-O bond length signifies that it is well activated by
Pt atom and O$_{vac}$ . Next, in step iii, CO molecule adsorbed on
Pt$_{1}$ atom with binding energy 1.27 eV, and one of Oxygen atom
of O$_{2}$ molecule comes nearer to CO molecule and form transition
state (TS-1) . The activation energy needed to process the reaction
($CO_{ad}+O-O_{ad}\rightarrow CO_{2}+O_{ad}$) is 0.49 eV, and after
releasing first CO$_{2}$molecule from remaining O$_{ad}$ atom restores
the Pt-loaded stoichiometric hematite surface in step iv. In step
v, another CO molecule adsorbed at Pt atom and migrated to an O$_{ad}$
atom in step vi and form a second transition state (TS-2). The activation
energy needed for the processing of the second reaction is 0.79 eV.
After releasing the second CO$_{2}$ molecule, the Pt-loaded stoichiometric
hematite surface reduced again to create new oxygen vacancy near Pt
atom and approaches to initial step i. All of the catalytic steps
are exothermic and the activation energy needed for formation of CO$_{2}$
molecule is small at low temperature, indicates that catalytic activity
of Pt$_{1}$/FeO$_{x}$ for CO oxidation is very high. 

Liang $et$ $al.$ and Qiao $et$ $al.$ investigated experimentally
as well as theoretically the catalytic activity of Ir$_{1}$/FeO$_{x}$\cite{Ir1/FeOx_CO_liang2014theoretical},
Au$_{1}$/FeO$_{x}$\cite{Au1/FeOx} for CO oxidation, respectively. 

Using DFT, Liang $et$ $al.$ explored the catalytic activity of non-noble
metal single-atom catalyst (i.e., Ni$_{1}$/FeO$_{x}$\cite{Ni1/FeOx_CO_liang2016theoretical})
and also compared the catalytic activity of Pt$_{1}$/FeO$_{x}$,
Ir$_{1}$/FeO$_{x}$ and Ni$_{1}$/FeO$_{x}$ systematically. The
O$_{2}$ molecule adsorbed on the surface of these SACs in a different
manner, in the case Pt$_{1}$/FeO$_{x}$ and Ni$_{1}$/FeO$_{x}$
it adsorbed on top of Pt and Ni atoms, respectively, whereas in the
case and Ir$_{1}$/FeO$_{x}$ it adsorbs dissociatively, i.e., one
O atom on the top of Ir and another O atom occupy the oxygen vacancy
site. The activation energy needed for the formation of CO$_{2}$
(TS-1) in the case of Pt$_{1}$/FeO$_{x}$, Ir$_{1}$/FeO$_{x}$ and
Ni$_{1}$/FeO$_{x}$ catalysis is 0.49 eV, 0.59 eV and 0.75 eV, respectively,
whereas for the formation of second CO$_{2}$ (TS-2) the activation
barrier is 0.79 eV, 1.41 eV and 0.64 eV, respectively. The rate-determining
step for Ni/FeO$_{x}$ (0.75 eV) catalyst is lowest compared to Pt$_{1}$/FeO$_{x}$
(0.79 eV) and Ir$_{1}$/FeO$_{x}$ (1.41 eV) catalyst, suggest that
it exhibits the highest catalytic activity for CO oxidation compared
to others at room temperature. 

Using experimental and theoretical methods, Moses-DeBusk $et$ $al.$\cite{Pt/Al2O3}
examine the catalytic activity of single Pt atom dispersed on an inert
substrate, $\theta-$Al$_{2}$O$_{3}$ for CO oxidation, in the presence
of stoichiometric oxygen. They reported that the proposed pathway
of CO oxidation is slightly different from the conventional Langmuir-Hinshelwood
mechanism because the conventional mechanism requires at least one
Pt-Pt bond. 

In search of non-precious and more efficient/active SACs for CO oxidation
Li $et$ $al.$\cite{M1/FeOx_li2014exploration} systematically studied
the catalytic activity of various single-atom catalysts M$_{1}$/FeO$_{x}$
(M=Au, Rh, Pd, Co, Cu, Ru and Ti) by employing density functional
theory. They reported five SACs, especially Rh$_{1}$/FeO$_{x}$and
Pd$_{1}$/FeO$_{x}$ with oxygen vacancy, CO$_{1}$/FeO$_{x}$and
Ti$_{1}$/FeO$_{x}$ without oxygen vacancy, and Ru$_{1}$/FeO$_{x}$
with or without oxygen vacancy surface exhibits better catalytic activity
compared to Pt$_{1}$/FeO$_{x}$. Furthermore, they also reported
that non-precious single atom CO$_{1}$/FeO$_{x}$and Ti$_{1}$/FeO$_{x}$
catalyst need very low activation energy for CO oxidation via L-H
mechanism.

Using DFT calculation Tang $et$ $al.$\cite{Pt/CeO2_tang2017theoretical}
systematically studied the catalytic activity of single Pt atom dispersed
on the CeO$_{2}$ (111), (110) and (100) surfaces for CO oxidation
via Mars-van Krevelen mechanism. They reported that the single Pt
atom loaded on the ceria surfaces are thermodynamically stable, and
the oxidation state of Pt atom on (111) and (100) surfaces is +4.
In contrast, the oxidation state of (110) surface is +2 due to the
spontaneous formation of O$_{2}{}^{2-}$ spices, which reduces the
oxidation state of Pt atom from +4 to +2, making the Pt$_{1}$@CeO$_{2}$
(110) catalyst most stable. 

\subsection{Water-Gas Shift Reaction }

The water-gas shift (WGS) reaction was discovered in 1780 by Italian
physicist Felice Fontana, but its importance in the industrial sector
was realized much later. In this reaction, carbon monoxide and water
vapor reacts to form carbon dioxide and hydrogen molecule $(CO+H_{2}O\rightleftharpoons CO_{2}+H_{2})$.
WGS is a cost-effective and more efficient method for the production
of hydrogen. In industrial sectors, a large amount of hydrogen is
needed for various process such as ammonia synthesis via Haber-Bosch
process, synthetic liquid fuels synthesis via Fischer-Tropsch method,
hydro-treating of petroleum products for removing CO contamination,
in the synthesis of nitrogenous fertilizers, for preparation of ethanol,
methanol, and dimethyl ether, and hydrogenation of hazardous wastes
(PCBs and dioxins)\cite{H_app_ramachandran1998overview,WGS_aap_Ratnasamy}. 

Apart from this, from future aspects, hydrogen is considered to be
one of the cleanest and renewable energy source because it can be
stored and transported efficiently and after burning, it produces
only water as a byproduct\cite{Hydrogen_pro_chen2010semiconductor,Hydrogen_Pro_levalley2014progress,Hydrogen_pro_pagliaro2010solar,Hydrogen_pro_Turner972}. 

Due to the high catalytic activity, selectivity, and efficiency of
SACs, many researchers have investigated its catalytic properties
for WGS reaction\cite{Ir1/FeOx,WGS_review_flytzani2012atomically,wgs_thomas2011can,WGS_Au_flytzani2013gold,Au/CeO2_WGS_DFTsong2014mechanistic,Au-OHx/TiO2_WGSyang2013atomically}.
Fu $et$ $al.$ \cite{Au/Ceo_WGS_fu2005activity} synthesized low-content
(0.2 - 0.9 wt\%) gold-cerium oxide catalyst and reported its activity
and stability is high for WGS reaction. Yang $et$ $al.$ prepared
SAC consisting of isolated Au atoms dispersed on titania support and
reported that it exhibits excellent activity for WGS reaction at low
temperatures. They stabilize Au atom on support by irradiating titania
support by UV rays, which is suspended in ethanol solution, where
the gold atom donates the separated electrons to $-$OH groups. The
Au atoms with surrounding extra surface $-$OH groups act as active
sites for the WGS reaction and also reported that its catalytic performance
is better than Au/CeO$_{2}$\cite{Au/Ceo_WGS_fu2005activity,Au/Pt/ceria_WGSfu2003active}.
Flytzani-Stephanopoulos group members prepared single atom centric
Pt (Pt(II)$-$O(OH)$_{x}-$) and Au (Au$-$O(OH)$_{x}-$) sites stabilize
by sodium or potassium ion by making bond with it through $-$O ligands
on three different supports, and examine its catalytic activity for
WGS reaction \cite{Pt(II)--O(OH)x_WGS_yang2015common,Au-O(OH)x_WGS_yang2014catalytically}.
They found that the reaction rate of Pt(II)$-$O(OH)$_{x}-$ species
for WGS reaction is same for all supports (i.e., anatase (TiO$_{2}$),
a microporous K-type L-zeolite (KTLZ) and mesoporous silica MCM-41
({[}Si{]}MCM41) ) for Na-containing catalyst with 0.5 wt\% Pt loading\cite{Pt(II)--O(OH)x_WGS_yang2015common}.
Similar to finding of single-site Pt(II)$-$O(OH)$_{x}-$ species,
irrespective of the support KLTL and {[}Si{]}MCM41, TiO$_{2}$, CeO$_{2}$,
and Fe$_{2}$O$_{3}$ the reaction rate of Au$-$O(OH)$_{x}-$ species
is same with 0.25 wt\%, 0.25 wt\%, 0.12 wt\%, 0.50 wt\% and 1.16 wt\%
Au loading, respectively\cite{Au-O(OH)x_WGS_yang2014catalytically}. 

Lin $et$ $al.$\cite{Ir1/FeOx} synthesized catalyst consisting of
isolated Ir atom loaded on FeO$_{x}$ support and found that it shows
remarkable performance for WGS reaction. The catalytic activity of
Ir$_{1}$/FeO$_{x}$ is higher than its cluster and nano-particle
counterparts, also higher than Au- or Pt-based catalyst\cite{Au/Pt/ceria_WGSfu2003active}.
After extensive research, they found that the single atom is responsible
for $\approx$70\% catalytic activity in a single atom, clusters and
nano-particles catalyst. The Ir atom helps FeO$_{x}$ support in reduction
for creating oxygen vacancy, which leads to enhance the catalytic
activity of Ir$_{1}$/FeO$_{x}$.

In literature, it has been seen that the WGS reaction mainly follows
three reaction mechanisms, i.e., redox, formate, and carboxyl mechanisms.
Fu et al.\cite{Au/Pt/ceria_WGSfu2003active} proposed that nano-structured
gold-ceria oxide catalyst follows the redox mechanism for WGS reaction.
In this reaction mechanism, CO atom adsorbed on Au atom and oxidized
with the help of O atom of ceria oxide; after that, support is reoxidized
by water, and hydrogen is released. Shido and Iwasawa\cite{Formate_WGW_SHIDO1992493,Formate_WGW_SHIDO199371}
were the first to propose the formate mechanism for WGS reaction,
in this mechanism CO and H$_{2}$O molecule adsorbs on the surface
of support, next CO molecule interact with surface OH group to form
formate (HCOO) intermediate, which dissociate into CO$_{2}$ molecule
and H atom, and finally, two H atom recombine to form H$_{2}$ molecule.
Liu et al.\cite{Carboxyl_PhysRevLett.94.196102}, studied the catalytic
activity Au clusters-ceria oxide (Au$_{4-6}$/CeO$_{2}$) catalyst
for WGS reaction and found that it follows the carboxyl mechanism.
In this mechanism, CO$_{ad}$ adsorbs on Au, and H$_{2}$O dissociatively
(H and OH) adsorbs on Au, the adsorbed CO$_{ad}$ interact with OH$_{ad}$
to form carboxyl (COOH) intermediate after that COOH dissociate into
CO$_{2}$ molecule and H atom, and at last, two H atom recombine to
form H$_{2}$ molecule. Song $et$ $al.$\cite{Au/CeO2_WGS_DFTsong2014mechanistic}
predicted the reaction mechanism of isolated and clustered Au atoms
on CeO$_{2}$(110) using density functional theory for WGS reaction. 

Song $et$ $al.$\cite{Au/CeO2_WGS_DFTsong2014mechanistic} by employing
DFT methods studied reaction mechanism of isolated and clustered Au
atoms on CeO$_{2}$ (110) surface for WGS activity, using both pathways
redox and carboxyl mechanism. The carboxyl mechanism is more favorable
than redox mechanism because it requires higher energy for breaking
O$-$H bonds, which is directly involved in the production of H$_{2}$
and CO$_{2}$.

\begin{figure}
\begin{centering}
\includegraphics[scale=0.4]{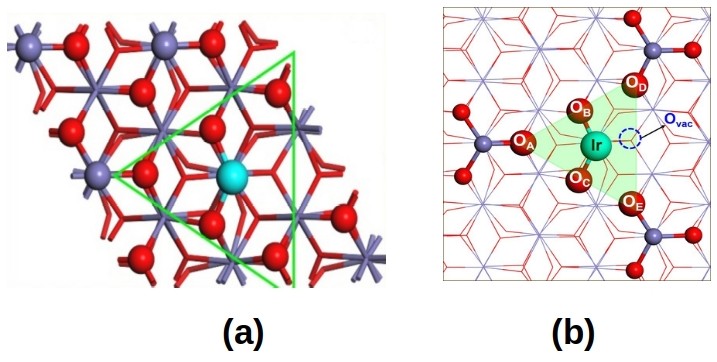}
\par\end{centering}
\caption{\label{fig:Ir1FeOx}(a) The top view of SAC Ir$_{1}$/FeO$_{x}$with
oxygen vacancy (O$_{vac}$) and (b) the local atomic arrangement of
Ir$_{1}$/FeO$_{x}$$-$O$_{vac}$ is shown in the figure. The surface
lattice oxygen atom of FeO$_{x}$ support are represented by O$_{A}$,
O$_{B}$, O$_{c}$, O$_{D}$, O$_{E}$, and oxygen vacancy (O$_{vac}$)
is situated at the right side of Ir atom. Ir atom= blue, Oxygen atom=
red, and Fe atom= purple. (reprinted/reproduce from Ref. {[}42{]}
with the permission of Wiley-VCH group, copyright 2020).}
\end{figure}

Recently, Liang $et$ $al.$\cite{liang2020dual} studied the catalytic
activity of Ir$_{1}$/FeO$_{x}$ SAC for WGS reaction by using theoretical
and experimental methods. In Fig. \ref{fig:Ir1FeOx} (a) and (b),
a schematic diagram 

\begin{figure}[H]
\begin{centering}
\includegraphics[scale=0.45]{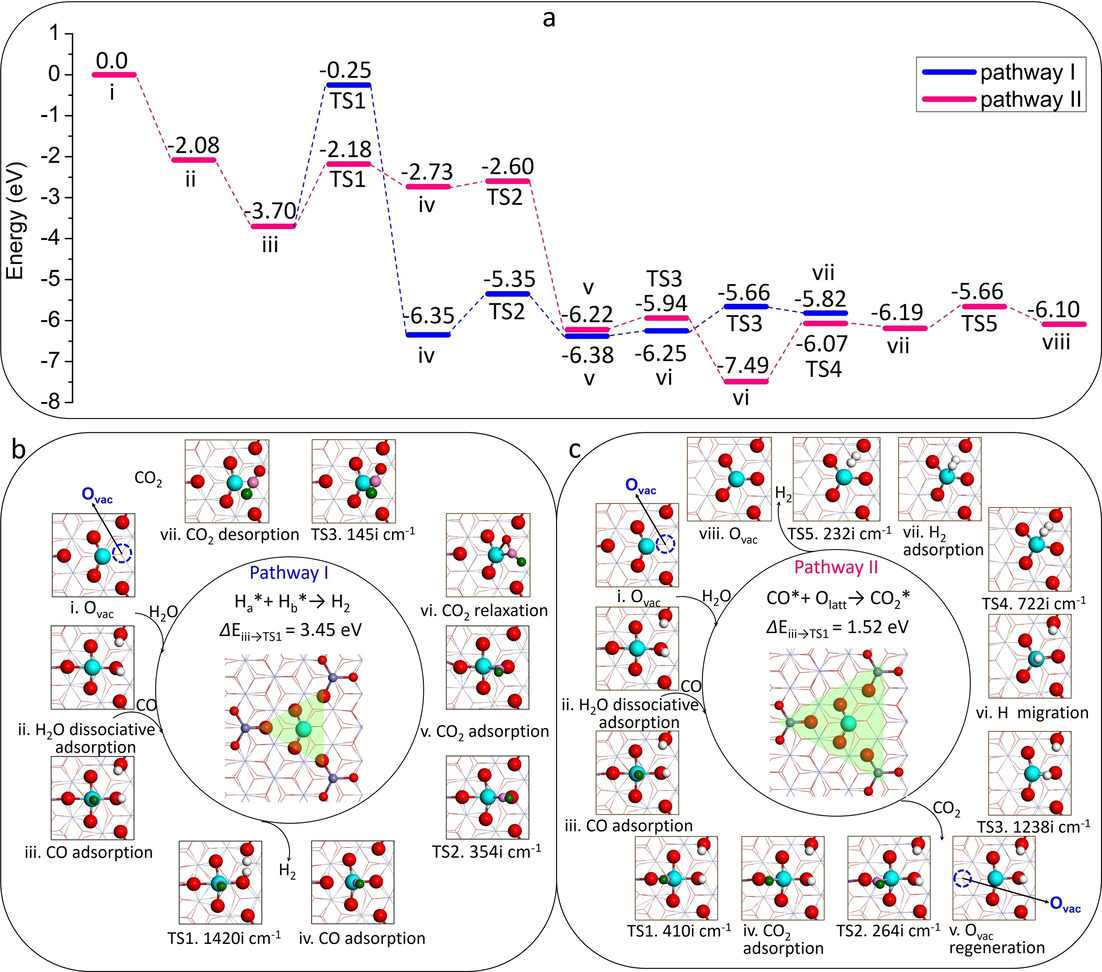}
\par\end{centering}
\caption{\label{fig:WGS_reaction_pathways}(a) The calculated relative energy
diagram of proposed reaction pathways I and II for WGS reaction on
Ir$_{1}$/FeO $_{x}$ catalyst. The reaction step and corresponding
structures for (b) path I and (c) path II are shown. The rate-determining
step with the energy barrier is demonstrated in a circle, and a green
triangle region represents active sites in the reaction. Ir atom=
blue, Oxygen atom= red, Fe atom= purple, C atom = pink and O atom
of CO= dark green. (reprinted/reproduce from Ref. {[}42{]} with the
permission of Wiley-VCH group, copyright 2020).}
\end{figure}

of Ir$_{1}$/FeO$_{x}$ with oxygen vacancy, and the surface lattice
oxygen atom (red) around Ir atom blue are represented, respectively.
The most favorable position of Ir atom to stabilize on the surface
of FeO$_{x}$ is O$_{3}-$terminated surface, where Ir atom is bonded
with three oxygen atom. The site structure of Ir$_{1}$/FeO$_{x}$
with and without oxygen vacancy is the same, and it follows two different
redox reaction pathways I and II, shown in Fig. \ref{fig:WGS_reaction_pathways}
for WGS reaction.

Let us considered the reaction pathways I (Fig \ref{fig:WGS_reaction_pathways}
(b)), in step (i) Ir atom is bonded with two oxygen, and on the right
side of the Ir atom, there is an O$_{vac}$. Next, in step (ii) H$_{2}$O
molecule dissociate into H and OH, and adsorbed on O$_{D}$ atom (represented
as H$_{a}$) and at O$_{vac}$ site (represented as O$_{F}$ for O
atom and H$_{b}$ for Hydrogen), respectively. The CO (O atom of CO
is represented as O$_{G}$) molecule is absorbed on Ir$_{1}$ atom
in step (iii). The next step is TS1, where H$_{a}$ and H$_{b}$ directly
combine to form H$_{2}$ and require high activation energy 3.45 $eV$,
which is also a rate-determining step. Afterwards the absorbed CO
atom starts moving towards the $O_{F}$ atom in step (iv) and gradually
approaches to TS2. The energy barrier for the formation of CO$_{2}$
if the activation energy of 1 $eV$ is applied. The newly form bent
CO$_{2}$ molecule with a 140.7$^{\circ}$ angle still absorbs on
Ir single atom in step (v). The bent CO$_{2}$ can be considered as
a virtual CO$^{-}{}_{2}$ anion, and its absorption energy on the
Ir$_{1}$/FeO$_{x}$ is 1.29 $eV$and require another intermediate
step (vi) for relaxation. The bent CO$^{-}{}_{2}$ transforms into
linear CO$_{2}$ by losing an electron in TS3, and the activation
energy of TS3 is 0.59 eV. Finally, in step (vii), the desorption of
CO$_{2}$ from the Ir$_{1}$/FeO$_{x}$ and regeneration of O$_{vac}$
occurs on the Ir atom's right side.

The redox reaction pathways II is presented in Fig. \ref{fig:WGS_reaction_pathways}
(c), and the reaction path till step (iii) is the same as pathways
I. The Next step is TS1, which needs activation energy of 1.52 $eV$
to move the adsorbed CO molecule slowly towards the adjacent O$_{a}$
at the left side of Ir atom. The bent CO$_{2}$ (CO$^{-}{}_{2}$ )
molecule is formed in step (iv). Afterward, the bent CO$_{2}$ requires
small activation energy 0.13 $eV$ for desorbing from the surface
Ir$_{1}$/FeO$_{x}$ by releasing an electron in TS2. The O$_{vac}$
on the left side of the Ir atom is produced after the desorption of
CO$_{2}$ in step (v). The H$_{b}$ atom of HO$_{F}$ starts moving
towards Ir atom slowly if small activation energy of 0.28 $eV$ is
applied in TS3. In the intermediate step (vi) H$_{a}$ and H$_{b}$
atoms are associated with O$_{D}$ and Ir atom, respectively. The
H$_{a}$ and H$_{b}$ approach towards each other in TS4 for the formation
H$_{2}$ (H$^{*}{}_{a}$ + H$^{*}{}_{b}$$\rightarrow$H$^{*}{}_{2}$),
and energy barrier for the reaction is 1.42 $eV$. The obtained H$_{2}$
slowly migrate towards the Ir atom in step (vii). Next, H$_{2}$ molecule
desorbs from the Ir$_{1}$/FeO$_{x}$ in TS5 with an energy barrier
of 0.53 $eV$. Finally, after releasing the H$_{2}$ molecule, the
surface of Ir$_{1}$/FeO$_{x}$ SAC is recovered, and O$_{vac}$ is
generated on the left side of Ir atom. During the WGS reaction process
on Ir$_{1}$/FeO$_{x}$ surface, O$_{vac}$ shifts from the right
side to the left side of Ir atom. 

On comparing the pathways, we found that H$_{2}$ is first formed
before CO$_{2}$ in pathways I, whereas in path II, it is reversed.
The most favorable pathways for WGS reaction on Ir$_{1}$/FeO$_{x}$
surface is path II, as the energy barrier for the rate-determining
step is 1.52 $eV$, much lower compared to the Path I (3.45 $eV$).
Using Bader charge analysis Liang $et$ $al.$\cite{liang2020dual}
also reported the oxidation state of Ir and Fe atom for both pathways.
In pathways I, only Ir atom changes its oxidation state, whereas,
in pathways II, both Ir and Fe atom changes oxidation state. In step
(i) of pathways II, the oxidation state of Ir and Fe$^{(a)}$ atom
are +3 and +2, respectively, whereas in the final step (viii) the
oxidation of Ir atom decreases from +3 to +2 and the oxidation state
of Fe atom increases from +2 to +3. We can conclude that in WGS reaction
Pathways II both Ir and Fe atom changes its oxidation state.

\section{\label{sec:Summary-and-Conclusions}Summary and Conclusions }

In this review article, we presented the recent advancement in the
field of single-atom catalysis with a focus on the various synthesis
methods and their application in varoius catalytic reactions, such
as CO oxidation\cite{Pt/FeOx,Ir1/FeOx_CO_liang2014theoretical,Ni1/FeOx_CO_liang2016theoretical,Au1/FeOx,Pt/Al2O3,M1/FeOx_li2014exploration,Pt/CeO2_CO_oxidation_ALD_wang},
water\textminus gas shift (WGS)\cite{Ir1/FeOx,WGS_review_flytzani2012atomically,wgs_thomas2011can,WGS_Au_flytzani2013gold,Au/CeO2_WGS_DFTsong2014mechanistic,Au-OHx/TiO2_WGSyang2013atomically,liang2020dual},
 etc. We also discussed the reaction mechanism of a single-atom catalyst
for different catalytic reactions from theoretical aspects using density
functional theory.

\section*{Acknowledgements}

D.K.R would like to thank Prof. Jun Li and Prof. Yang-Gang Wang for useful discussion and for giving the opportunity to write a review article on Single-Atom Catalysts. The author also gratefully acknowledges the financial support from the Southern University of Science and Technology (SUSTech) and computational resource support from the Center for Computational Science and Engineering at SUSTech.

\bibliography{SAC}

\end{document}